\documentstyle[aps,psfig,twocolumn,floats,prb]{revtex}
\begin{document}

\def\gtwid{\mathrel{\raise.3ex\hbox{$>$\kern-.75em\lower1ex\hbox{$\sim$}}}}
\def\ltwid{\mathrel{\raise.3ex\hbox{$<$\kern-.75em\lower1ex\hbox{$\sim$}}}}
\def\ed{{\sl edited\ by\ }}
\def\cf{{\sl cf.~}}
\def\viz{{\sl viz.~}}
\def\etal{{\sl et.~al.}}
\def\tj{$t$-$J$~}

\draft
\flushbottom
\twocolumn[
\hsize\textwidth\columnwidth\hsize\csname @twocolumnfalse\endcsname

\title{Truncated Lanczos method~: application to hole-doped spin ladders}

\author{O.~Chiappa, S.~Capponi and D.~Poilblanc}
\address{
Laboratoire de Physique Quantique \& UMR--CNRS 5626,
Universit\'e Paul Sabatier \\
F-31062 Toulouse, France}

\date{\today}
\maketitle
\tightenlines
\widetext
\advance\leftskip by 57pt
\advance\rightskip by 57pt

\begin{abstract}
The truncated Lanczos method using a variational scheme based on Hilbert
space reduction as well as a local basis change (Dagotto {\it et al.},
Phys.~Rev.~B~{\bf 58}, 12063 (1998)) is re-examined . The energy is
extrapolated as a power law function of the Hamiltonian variance. This
systematic extrapolation procedure is tested quantitatively on the two-leg
\tj ladder doped with two holes. For this purpose, we have carried out calculations of the spin gap and of the pair dispersion up to size $2 \times 15$.
\end{abstract}
\vskip2pc]
\narrowtext
\tightenlines

The  variational approach of the truncated Lanczos method (TLM) \cite{Rie93,Dag98} is based first on a proper choice of the local basis, and then on Hilbert space reduction and exact diagonalizations by the Lanczos algorithm in the successive subspaces. This variational technique is a good alternative to the density matrix renormalization group (DMRG) calculations \cite{dmrg} and the pure exact diagonalization technique (ED) for the study of models related to high temperature superconductivity. Another well known method to study numerically the low energy physics of fermionic models on medium size clusters is the Quantum Monte Carlo algorithm (QMC)\cite{QMC:1} hindered by the famous sign problem. On the other hand, the DMRG calculations provide only properties of the lowest energy sector and operate on lattice systems with specific boundary conditions. The TLM  goes beyond these technical limitations and allows us to use the ED technique on larger fermionic systems.
This approach is very useful e.g. to study momentum-dependence of the observables and calculate dynamical properties of Hubbard-like models or \tj model on large size systems (compared to standart ED techniques) of various geometries. The issue we shall adress here is the control of the convergence of the energy. For this purpose, we will exploit an interesting behaviour of the energy with the Hamiltonian variance, defined as the measure of the fluctuations of the energy around its average. We will show that the energy extrapolates with a second order power law function with the variance \cite{Sor:1}, such that the extrapolations to the zero-variance limit are accurate estimations of the approximate variational energy in a specified sector of symmetry. This unbiased extrapolating procedure provides more control on the accuracy of the results. The applications to the $2\times L$ hole-doped spin ladder\cite{Rice} (for which many exact calculations can be performed up to $L=13$) show that the error made on the exact values are small compared to the degree of approximation made. 
This article begins first with a detailed description of the method, followed by a discussion on the efficiency of the TLM through applications to the two-hole doped \tj model on the $2\times L$ ladder geometry with periodic boundary conditions for the sizes $L=13$, which can be handled exactly, and $L=15$.

%
%
\paragraph{Geometry:}

We will consider closed systems that can be formally described like a chain of a finite number of interacting finite size sub-systems or clusters.
Each cluster ${\cal S}^{\rm i}$ is described by a numerically soluble cluster Hamiltonian ${\sf H}_0^{\rm i}$ and two near-neighbor clusters ${\cal S}^{\rm i}$ and ${\cal S}^{\rm j}$ are interacting through the transverse Hamiltonian ${\sf H}_\perp^{\rm ij}$. The cluster Hamiltonian operator ${\sf H}_0^{\rm i}$ acts on the ${\rm i}^{\rm th}$-cluster Hilbert space ${\cal E}_0^{\rm i}$ while the transverse Hamiltonian operator ${\sf H}_\perp^{\rm ij}$ acts on the direct product ${\cal E}_0^{\rm i}\otimes{\cal E}_0^{\rm j}$.
Typically, ${\sf H}_\perp^{\rm ij}$ contains Heisenberg (magnetic couplings) and kinetic interaction terms (one-electron hopping integral). 
For example, in two chain ladders, that we will consider in the following, ${\cal S}^{\rm i}$ is simply a rung with two lattice sites.
The Hamiltonian of the global system is thus a sum of two terms
 ${\sf H}={\sf H}_0+{\sf H}_\perp$ with
${\sf H}_0=\bigoplus_{{\rm i}=1,{\cal N}}{\sf H}_0^{\rm i}$
and ${\sf H}_\perp=\sum_{<{\rm i,j}>}{\sf H}_\perp^{\rm ij}$
, ${\cal N}$ being the number of clusters constituting the whole system.

\paragraph{Change of basis:}

We begin with the exact diagonalization of ${\sf H}_0^{\rm i}$ of all the clusters ${\cal S}^{\rm i}$. The eigenstates of the ${\rm i}^{\rm th}$-cluster Hamiltonian ${\sf H}_0^{\rm i}$ are classified according to their quantum number (filling, transverse and longitudinal momentum, ${\rm S}_{z}$ spin component).
Let 
${\cal B}_{0}^{\rm i}=\{\vert\phi_{\rm p}^{\rm i}>\}_{\,{\rm p}=1,...,{\cal P}_{0}}$ 
with ${\cal P}_{0}={\rm dim}\,{\cal E}_{0}^{\rm i}\,$
denote the local basis set of the eigenstates of 
${\sf H}_0^{\rm i}$ in ${\cal E}_0^{\rm i}$. 
The direct product 
${\cal E}_0={\cal E}_0^{1}\otimes{\cal E}_0^{2}\otimes\,...\,\otimes{\cal E}_0^{\cal N}$ 
of the ${\cal N}$ cluster Hilbert spaces ${\cal E}_0^{\rm i}$ defines the Hilbert space of the whole system.
The initial basis set ${\cal B}_{0}$ of ${\cal E}_0$ in a given sector of symmetry is thus defined by the following set of ${\cal N}$-tensorial product vectors
$\Bigl\{
\vert\Phi_{\rm q}>=
\bigotimes_{{\rm i}=1,{\cal N}}\vert\phi_{{\rm p}_{\rm i}}^{\rm i}>
\Bigr\}
$ reduced to the states consistent with the quantum numbers implemented for the global system (in practice ${\cal Q}_{0}={\rm card}\,{\cal B}_{0} \ll {\cal P}_{0}^{\cal N}$).
Typically, use of translational symmetry in the longitudinal direction provides a significant reduction of ${\cal B}_{0}$.
We notice that ${\sf H}_0$ is naturally diagonal in ${\cal B}_0$. Its eigenvalues are just the sum 
${\rm h}_{{0},{\{{\rm p}_{1},...,{\rm p}_{\cal N}\}}}={\rm h}^{1}_{{0},{{\rm p}_{1}}}\,+\,{\rm h}^{2}_{{0},{{\rm p}_{2}}}\,+\,...\,+\,{\rm h}^{\cal N}_{{0},{{\rm p}_{\cal N}}}$ 
where the set $\{{\rm p}_{1},...,{\rm p}_{\cal N}\}$ is labelling a particular set of cluster Hamiltonian's eigenstates
(${\rm h}^{\rm i}_{{0},{\rm p}_{\rm i}}$ 
is the eigenvalue of ${\sf H}_0^{\rm i}$ associated to 
$\vert\phi_{{\rm p}_{\rm i}}^{\rm i}>$).

\paragraph{Expansion and truncation of the Hilbert spaces:}

\begin{figure}[h]
\begin{center}
\psfig{figure=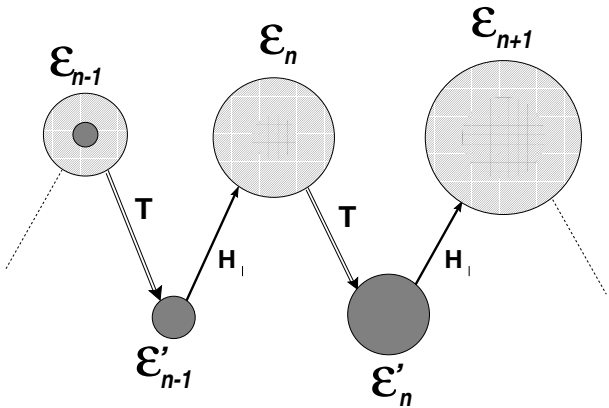,width=8truecm,angle=0}
\end{center}
\vspace{-0.2truecm}
\caption{Schematic picture of the successive Hilbert space truncation and expansion procedures. The operator ${\cal T}$  acts on the Hilbert spaces ${\cal E}_{n}$ to obtain the reduced spaces ${\cal E}^\prime_{n}$. We described precisely in the text the action of the non-unitary operator ${\cal T}$. 
}
\label{Fig:02}
\end{figure}

We have represented on Fig.~\ref{Fig:02} a picture of the iterative scheme.
Although ${\cal B}_{0}$ can be formally constructed it would be impossible to diagonalize exactly the Hamiltonian in this basis for sizes of interest, so that a truncation procedure is necessary. Keeping a small identical subset of the states in all the local basis ${\cal B}_{0}^{\rm i}$ would be in fact a quite poor approximation and we shall use a better procedure.
The first step consists in choosing a subset ${\cal B}_{0}^\prime$ of ${\cal B}_{0}$.
A simple choice is to restrict ourselves to the few low energy states of ${\cal B}_{0}^{\rm i}$.
The exact diagonalization by the Lanczos algorithm of the Hamiltonian ${\sf H}={\sf H}_0+{\sf H}_\perp$ in ${\cal B}_{0}^\prime$ gives the energy ${\rm e}_{0}^\prime$ in a fixed sector of symmetry at the ${\rm zero}^{\rm th}$-order of the iterative expansion. The ${\rm zero}^{\rm th}$-order approximation of the ground state is then a normalized linear combination of the vectors in ${\cal B}_{0}^\prime$.
At the first order, the expanded variational basis ${\cal B}_{1}$ is calculated by applying the transverse part ${\sf H}_\perp$ on the states in ${\cal B}_{0}^\prime$ $\left({\rm card}\,{\cal B}_1\,\gg\,{\rm card}\,{\cal B}_{0}^\prime \right)$. Like previously, we calculate the first order of the variational energy ${\rm e}_{1}$ by an exact diagonalization of ${\sf H}$ in the Hilbert space ${\cal E}_{1}$ generated by ${\cal B}_{1}$ (of course ${\rm e}_{1}<{\rm e}_{0}^\prime$).
Then the best states (see later) in ${\cal B}_{1}$ which constitute the new reduced basis ${\cal B}_{1}^\prime$ are selected. The reduction of the Hilbert space ${\cal E}_{1}$ ends the first iteration. The rest of the calculation has the same loop-structure. Let us briefly recall the scheme of the ${\left({\rm n-1}\right)}^{\rm th}$ iteration (Fig.~\ref{Fig:02}). It consists in two steps~:
$\left({\it i}\right)$  Expanding ${\cal B}_{n-1}^\prime$ by applying ${\sf H}_\perp$. Let ${\cal B}_{n}$ denote the new expanded basis $\left({\rm card}\,{\cal B}_{n}\,\gg\,{\rm card}\,{\cal B}_{n-1}^\prime\right)$, $\left({\it ii}\right)$ Truncating the generated basis ${\cal B}_{n}$, which give the reduced basis ${\cal B}_{n}^\prime$.

The normalized variational eigenvector $\vert\Psi_{n}>$ at order n of the truncation procedure is a linear combination of the ${\cal Q}_{n}={\rm dim}\,{\cal E}_{n}$ vectors of the subset ${\cal B}_{n}$ of ${\cal B}_{0}$.
In order to reduce the dimension of the Hilbert space ${\cal E}_n$ $\left(\,n\geq 1\,\right)$ we use a systematic procedure to select the best states in $\vert\Psi_{n}>$. 
Let ${\cal T}$ denotes the non-unitary truncation operator defined by the following Hilbert spaces morphism~:
\begin{eqnarray}
\nonumber
{\cal T}\,:\quad{\cal E}_{n}& \longrightarrow& {\cal E}_{n}^\prime\\
\vert\Psi_{n}>&
\longrightarrow&
\vert{\cal T}\Psi_{n}>\,=\,\sum_{{\rm q}\in{{\rm I}_n}^{\prime}}\alpha^{(n)}_{\rm q}\,\vert\Phi_{\rm q}>
\end{eqnarray}
where ${{\rm I}_n}^\prime$ is the set containing the indexes of the basis states $\vert\Phi_{\rm q}>$ that belong to the reduced basis ${\cal B}_{n}^\prime$.
The DMRG method also use such approximations. 
In order to select the best states we define a parameter $\epsilon\,\in\,[0,1]$, such that a state $\vert\Phi_{\rm q}>$ belongs to the reduced basis if the associated weight obeys to $\vert\vert\,\alpha^{(n)}_{\rm q}\,\vert\vert\,^{2}\geq\epsilon$. The actual value of $\epsilon$ at each iterative step can be determined by three different methods~:
$\left({\it i}\right)$  by specifying the minimum value of the overlap ${\cal O}\,=\,<\Psi_{n}\vert{\cal T}\Psi_{n}>\,=\,\sum_{{\rm q}\,\in\,{{\rm I}_n}^\prime}\vert\vert\,\alpha^{(n)}_{\rm q}\,\vert\vert\,^2$, $\left({\it ii}\right)$ by fixing the fraction ${\eta}$ of the states of ${\cal B}_n$ kept in the reduced basis set ${\cal B}_n^\prime$ and $\left({\it iii}\right)$ by fixing an expansion parameter $\lambda\,>\,1$ such that ${\rm dim}\,{\cal E}^\prime_{n}\,=\,\lambda\cdot{\rm dim}\,{\cal E}^\prime_{n-1}$.
In the following, procedure $\left({\it iii}\right)$ is used (for $n\,\geq\,2$) since it gives more control on the size growth of the Hamiltonian matrix and on the CPU time of the calculation.
The first reduced space ${\cal E}^\prime_1$ is obtained by using the criterium of minimum overlap, ie $<\Psi_{1}\vert{\cal T}\Psi_{1}>={\cal O}_{1}\,\geq\,{\cal O}_{\rm min}$. Note that for moderate expansion parameter $\lambda$ (see table~\ref{dim_1}), a rather large fraction of the weight is preserved at each step.
\begin{table}[h]
\begin{tabular}{c|ccc|ccc}
&\multicolumn{3}{c|}{$\lambda\,=\,2.27$}&\multicolumn{3}{c}{$\lambda\,=\,1.51$}\\
${\rm n}^{\rm th}\,{\rm loop}$&${\cal O}$&${\rm dim}\,{\cal E}^\prime_{n}$&${\rm dim}\,{\cal E}_{n}$&${\cal O}$&${\rm dim}\,{\cal E}^\prime_{n}$&${\rm dim}\,{\cal E}_{n}$\\ \hline
0&-&2 048&33 804 460&-&2048&33 804 460\\
1&0.9990  &10 278    &21 504 &0.9991 &10 278 &21 504\\
2&0.9967 &23 299    &83 679 &0.9927 &15 476  &83 679\\
3&0.9959 &52 939   &169 045   &0.9869& 23 320 &126 031\\
4&0.9961 &119 996  &327 084  &0.9812& 35 099& 174 257\\
5&-&-&-& 0.9761  &52 846 &245 066\\
6&-&-&-& 0.9772 &79 599  &347 059\\
7&-&-&-& 0.9826  &119 967 &472 799
\end{tabular}
\caption{Evolution of the Hilbert space dimensions with the truncation criteria $\lambda\,=\,2.27$ and $\lambda\,=\,1.51$, for the $2\times13$ two-hole doped spin ladder in the $\left(S_z=0,\,{\mathbf K}=(0,0)\right)$ sector. We reported also in the first column the overlap ${\cal O}$ (defined in the text).
The basis ${\cal B}_{0}^\prime$ is built with the $2048$ lowest energy eigenstates of ${\sf H}_{0}$ in  ${\cal B}_{0}$ (with two holes on the same rung and particles on the other sites) compatible with the global quantum numbers. }
\label{dim_1}
\end{table}

\paragraph{Extrapolation procedure:}

As it will be clear in the following, it is very useful to define the Hamiltonian variance $\sigma_n$ at the ${\rm n}^{\rm th}$ iteration :
\begin{equation}
\sigma_n\,=\,{1\over {N^{2}}}\,
\{
{<\Psi_{n}^\prime\vert\,{\sf H}\,^2\,\vert\Psi_{n}^\prime>-<\Psi_{n}^\prime\vert\,{\sf H}\,\vert\Psi_{n}^\prime>^2}
\}
\end{equation}
where $N$ is the number of sites on the lattice and the new wave function $\vert\Psi_{n}^\prime>$ is the normalized eigenstate of ${\sf H}$ in the truncated basis ${\cal B}_{n}^\prime$ with the lowest eigenenergy.
The calculation of $\sigma_n$ does not need any approximation and can be exactly evaluated. 
$<\Psi_{n}^\prime\vert\,{\sf H}\,^2\,\vert\Psi_{n}^\prime>$ is calculated by expressing the vector ${\sf H}\,\vert\Psi_{n}^\prime>$ in ${\cal B}_{n+1}$ and evaluating its norm, while $<\Psi_{n}^\prime\vert\,{\sf H}\,\vert\Psi_{n}^\prime>^2$ is easily obtained by the exact diagonalization of ${\sf H}$ in ${\cal B}_{n}^\prime$.
$\sigma_n$ measures the quadratic deviations of the energy from its averaged value at the ${\rm n}^{\rm th}$ iteration. We  have tried four kinds of extrapolation of the energy versus the Hamiltonian variance :
$\left({\it i}\right) \, {\rm e}_{n} \, {\rm vs} \, \sigma_{n}$,
$\left({\it ii}\right) \, {\rm e}_{n}^\prime \, {\rm vs} \, \sigma_{n}$, 
$\left({\it iii}\right) \, {\rm e}_{n+1} \, {\rm vs} \, \sigma_{n}$
and
$\left({\it iv}\right) \, {\rm e}_{n+1}^\prime \, {\rm vs} \, \sigma_{n}$
where ${\rm e}_{n}\,=\,<\Psi_{n}\vert\,{\sf H}\,\vert\Psi_{n}>$,
with $\vert\Psi_{n}>$ the normalized eigenwave function associated to
the lowest eigenenergy of 
${\sf H}$ in ${\cal B}_{n}$, 
and ${\rm e}_{n}^\prime\,=\,<\Psi_{n}^\prime\vert\,{\sf H}\,\vert\Psi_{n}^\prime>$.
The zero-variance limits of the extrapolation curves are approximations of the exact energy. 
Note that each value of the variance corresponds in fact to a distribution of energies
since two variational states of same energies might have different
variances. Although we did not explicitely evaluate this distribution
of energy, this uncertainty is directly connected to the (estimated) difference
between ${\rm e}_{n}^\prime$ and  ${\rm e}_{n}$ for the same variance.
Hence, this gives a procedure to take into account this uncertainty in energy while 
extrapolating our results and we shall discuss quantitatively the precision of this
technique through applications 
to the \tj model defined on the two-leg ladder geometry. 

\begin{figure}[h]
\begin{center}
\psfig{figure=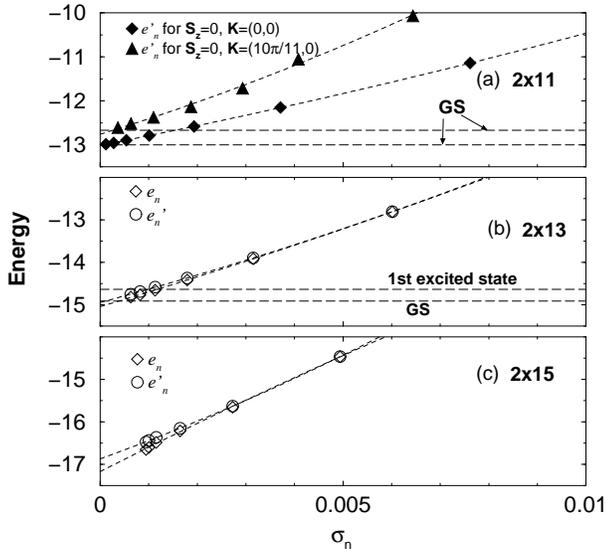,width=8truecm,angle=-90}
\end{center}
\vspace{-0.2truecm}
\caption{Energy of a $2\times 11$ (a), $2\times 13$ (b) and $2\times 15$ (c)
ladder with two holes extrapolated as a quadratic function of the
Hamiltonian variance compared to ED results when available. We made the calculations for 6 iterations with a final Hilbert size fixed to $120\,000$ symmetric states in the sector $\left(S_z=0,\,{\mathbf K}=(0,0)\right)$ for the (b) and (c) cases, while in (a) we plotted the results for 8 iterations with a final Hilbert space quite equal to the complete Hilbert space of the $2\times 11$ ladder with two holes ($1\,940\,064$ states without the spin inversion symmetry) in the sectors $\left(S_z=0,\,{\mathbf K}=(K_x,0)\right)$ with $K_x=0$ and $K_x\simeq\pi$.
}
\label{Fig:01}
\end{figure}
%
%
\paragraph{Applications to the $2\times L$ hole-doped ladder:}

\begin{figure}[h]
\begin{center}
\psfig{figure=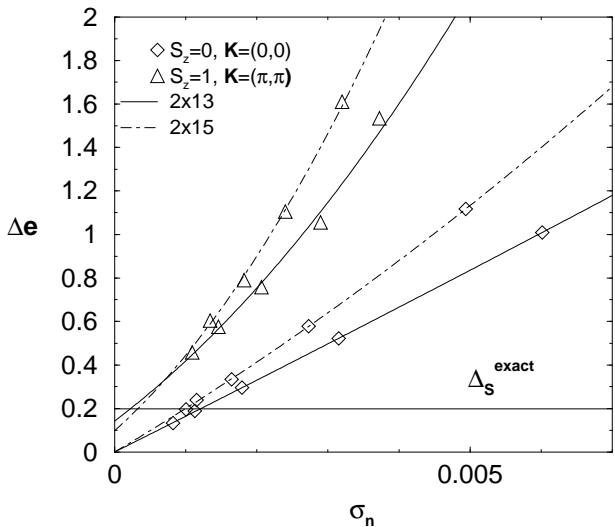,width=8truecm,angle=-90}
\end{center}
\vspace{-0.2truecm}
\caption{
Extrapolations of $\Delta e=(e_n+e'_n)/2-e_\infty$ where $e_\infty$ is the
zero-variance limit of the averaged GS energy 
for the $2\times 13$ and $2\times 15$ ladders with two holes in the sectors $\left(S_z=0,\,{\mathbf K}=(0,0)\right)$ and $\left(S_z=1,\,{\mathbf K}=(\pi,\pi)\right)$.
 We made the extrapolation of the energy calculated for 6 iterations with
$120\,000$ symmetric states in the last truncated Hilbert space ($0.37\%$
and $0.0021\%$ of the whole Hilbert space of the $2\times 13$ and $2\times
15$ ladders). $\Delta_S$ is obtained as the zero-variance extrapolation
of the upper curves and compared to ED result for $2\times 13$. 
}
\label{Fig:03}
\end{figure}

In this section, we apply the TLM to the \tj model on the $2\times{L}$ Heisenberg spin ladder with two holes and periodic boundary conditions. The Hamiltonian is defined by
\begin{eqnarray}
{\sf H}&=&J \sum_{i,\lambda} ({\mathbf S}_{i,\lambda}\cdot
{\mathbf S}_{i+1,\lambda}-\frac{1}{4}n_{i,\lambda}n_{i+1,\lambda}) \\
&+&J\sum_{i} ({\mathbf S}_{i,1}\cdot {\mathbf S}_{i,2}
-\frac{1}{4}n_{i,1}n_{i,2})\nonumber \\
\quad+&t& \sum_{i,\lambda,s}
(c_{i,\lambda,s}^{\dagger} c_{i+1,\lambda,s}^{\phantom{\dagger}}
+ h.c.)+t \sum_{i,s} (c_{i,1,s}^{\dagger}
c_{i,2,s}^{\phantom{\dagger}} +h.c.) \nonumber 
\label{one}
\end{eqnarray}
Here $c^\dagger_{i,\lambda, s}$ creates an electron of spin $s$ on site $i$
of leg $\lambda=1$ or 2, $ {\mathbf S}_{i,\lambda} = 
 (c^\dagger_{i,\lambda,s}
\  \sigma^{\phantom{\dagger}}_{ss^\prime} c^{\phantom{\dagger}}_{i,\lambda,s^\prime})/2$ 
and $n_{i,\lambda}= \Sigma_s c^\dagger_{i,\lambda,s}
c^{\phantom{\dagger}}_{i,\lambda,s}$.  We have taken both the nearest-neighbor leg and rung
one-electron hopping matrix elements equal to $t=-1$.
  The exchange interaction $J$
is taken as isotropic between nearest-neighbor leg and rung sites as well and
throughout $J/t=0.5$.
Extrapolations of the energy versus the Hamiltonian variance are shown on
Fig.~\ref{Fig:01} for the $2\times13$ and $2\times15$ two-hole doped ladder
in the sector $\left(S_z=0,\,{\mathbf K}=(0,0)\right)$. These extrapolations
are mainly linear, with a quadratic correction at low variance. The energies
corresponding to the zero-variance limit on these curves should be compared
to the exact ground-state (GS) ones ${\rm E}_{\rm GS}$ (for sizes $2\times11$ and $2\times13$ see Ref.~\cite{Note2}). 
For the $2\times13$ ladder, we noticed first that there is no significative difference between the extrapolations corresponding either to 5,6,7 and 8 iterations, but also for those associated to final Hilbert space sizes of $120\,000$ and $200\,000$ states.  
The stability of the extrapolations with variations of these parameters is an indication of a good precision on the approximate energy.
 The two best approximations of the variational energy come from the extrapolations of ${\rm e}_{n+1}$ and ${\rm e}_{n+1}^\prime\,{\rm vs}\,\sigma_n$. We minimize the error by averaging the two values of the energy obtained by these extrapolations. 
For the $2\times13$ system, this procedure leads to an error on the energy less than 
$1.85\%$ of the energy gap between the first excited level and the fundamental ($0.005$ compared to $\simeq0.27$), while we took into account only $0.37\%$ of the states in the Hilbert space of the whole system. This means that only a negligible fraction of the basis states in the fundamental wave function have a non-negligible contribution to the total weight.
The calculations on the $2\times15$ ladder with the same parameter
$\lambda$ generate an averaged extrapolation with more uncertainty
($\simeq 0.2$) than the one observed for smaller system sizes. 
This is mainly due to the decreasing of the overlap of the truncated
states at each 
step of the calculation, so that we should increase the dimension of
the final Hilbert 
size in order to compensate this loss in precision. Meanwhile, the
evolution of the error 
bars give us an information on the accuracy of the extrapolated values
of the variational 
energy, in such a way that we are able to adjust the initial
paramaters to reduce the 
error margin. 
We have attempted a systematic extrapolation procedure within this method.
As a test, we will complete calculations of the spin gap and of the pair dispersion of the low lying $S_z=0$ pair state for two-hole doped $2\times{L}$ ladders up to size ${L}=15$. 
As in a preceeding work on the evolution of the spin gap upon doping a 2-leg ladder \cite{Pb00} we defined the spin gap as $\Delta_S = E_0 \left(n_h=2, S=1\right) - E_0 \left(n_h=2, S=0\right)$. In this paper, we used the ED Lanczos method on $2\times{L}$ ladders of maximum size ${L}=13$. 
With a less important numerical effort, we expect improving the precision of the TLM by comparing the spin gap obtained with the TLM for ${L}=13$ with the exact calculations.
Our data follow in general a power law behaviour (Fig.~\ref{Fig:03}). Note that in the sector  $\left(S_z=0,\,{\mathbf K}=(K_x,0)\right)$ the extrapolations are dominated by the linear behaviour for low momentum while they are rather quadratic for higher momentum in the same sector and for $\left(S_z=1,\,{\mathbf K}=(\pi,\pi)\right)$.
Thus, the spin gap appears logically to be extrapolated by the
polynomial function
$\Delta_S=a_{0}+a_{1}\,\sigma_n+a_{2}\,\sigma_n^2$. The averaged
values in the zero-variance limit are $\Delta_S=0.14$ for ${L}=13$ and
$\Delta_S=0.10$ for size ${L}=15$, while the exact value is
${\Delta_S}^{\rm exact}=0.19$ for $L=13$. The lost in accuracy of $\Delta_S$
for ${L}=15$ is connected to the increased uncertainty of the average of the
extrapolated values due to the decrease of the proportion of states kept in the last 
truncated Hilbert space.

\begin{figure}[h]
\begin{center}
\psfig{figure=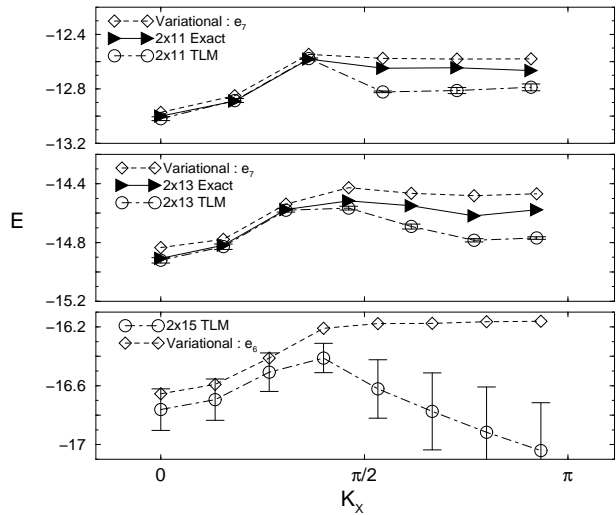,width=8truecm,angle=-90}
\end{center}
\vspace{-0.2truecm}
\caption{Pair dispersion calculated with the truncated Lanczos method on the $2\times 11$, $2\times 13$ and $2\times 15$ ladders with two holes in the sector $\left(S_z=0,\,K_y=0\right)$ (see the text for details).
}
\label{Fig:04}
\end{figure}

Lastly we apply this systematic extrapolation procedure to study the pair dispersion in the sector $\left(S_z=0,\,K_y=0\right)$. The results obtained within the TLM for ${L}=11$, ${L}=13$ and ${L}=15$ are shown on Fig.~\ref{Fig:04}. 
For the $2\times11$ and $2\times13$ ladders with two holes, we plot the variational energies ${\rm e}_{n}$ obtained at the $n=7^{\rm th}$ iteration for final truncated Hilbert space sizes fixed respectively to $120\,000$ and $1\,300\,000$ states, which correspond approximatively to the same proportion of the complete Hilbert space sizes ($6.5\%$ and $4\%$ respectively). In the case of the $2\times15$ ladder, we kept $120\,000$ states ($\simeq\,0.0021\%$ of the whole space) and we made 6 iterations. Let us recall that the error bars correspond to the difference between the extrapolations to the zero-variance limit of ${\rm e}_{n+1}$ and ${\rm e}_{n+1}^\prime$ in function of $\sigma_{n}$. The loss of accuracy on the averaged value of the extrapolated energies is obvious when the ladder size is increased for the same number of states in the last truncated Hilbert space, and we notice also that the error on the extrapolated energies for the $2\times 15$ ladder is also increasing from $0.2$ to $0.6$ for $K_x\,>\,\pi/2$. For small sizes, the extrapolated pair dispersion coincides with the exact one in the low momentum sector ($K_x\,\leq\,\pi/2$). For high momentum ($K_x\,>\,\pi/2$), we notice on the three figures a separation between the pair dispersion obtained with the variational energies which remains close to the exact one and the average of the extrapolated energies. This fact is probably related to the proximity of the continuum of the excitated states in high momentum sectors. The dispersion of energies for each value of the variance is increasing, so that our punctual extrapolations become less accurate than in the low momentum sector. Clearly this point needs to be more precisely elucidated.

To conclude, we have described and tested a truncation  procedure
which can be applied
to a large class of strongly correlated systems. This is a 
tool to construct variational approximations of the exact ground
state by only keeping a small fraction of the whole Hilbert
space. Here, we have focussed on the reliability of the method
to produce accurate energies, an issue not addressed in details 
in the litterature so far.
We have discovered an interesting behavior of the variational
energy as a function of the variance of its corresponding state which
enables to quantitatively estimate error bars. 
The method has been tested in the case of a two-hole doped two-leg
ladders and we have found that the accuracy depends strongly on 
details of the excitation spectrum (presence of a gap, of a continuum...).
In particular, the accuracy of the two-hole pair dispersion
is better for momenta where the ground-state is well separated
from the continuum. 
Although, the method is certainly not as accurate as e.g. the DMRG
technique, we believe it can offer a complementary analysis 
in case where the DMRG is less efficient (e.g. for systems with
periodic boundaries) while enabling to use symmetry analysis as in
standard exact diagonalisation methods.
Systems where part of the charge or spin fluctuations are selectively
suppressed by external potential (e.g. stripe potential, etc...) are
best candidates for this method. 

We thank  IDRIS (Orsay) for allocation of
CPU time on the NEC SX5.

\end{document}